\documentclass[amsmath, amssymb, preprintnumbers, showpacs, showkeys,aps,prl,superscriptaddress,twocolumn]{revtex4-2}

\usepackage{graphicx}
\usepackage{dcolumn}
\usepackage{bm,upgreek}
\usepackage{soul}
\usepackage[normalem]{ulem}
\usepackage{xcolor}
\definecolor{darkblue}{RGB}{0,0,139}

\usepackage{xcolor}
\usepackage{physics}
\usepackage{hyperref}
\hypersetup{
    colorlinks=true,
    linkcolor=blue,
    filecolor=magenta,      
    urlcolor=cyan,
    citecolor =darkblue
      }

\usepackage{bibunits}
\usepackage{printlen}
\uselengthunit{mm}

\begin{document}

\newcommand{\papertitle}{Polarization-controlled transport of light in a two-dimensional waveguide}
\title{ \papertitle}

\author{J.V.S. Ferreira}
\affiliation{Departamento de F\'{\i}sica, Universidade Federal de S\~{a}o Carlos, Rodovia Washington Lu\'{\i}s, km 235 - SP-310, 13565-905 S\~{a}o Carlos, SP, Brazil}
\affiliation{Universit\'e C\^ote d'Azur, CNRS, INPHYNI, France}

\author{A.M.G de Melo}
\affiliation{Universit\'e C\^ote d'Azur, CNRS, INPHYNI, France}

\author{N.A. Moreira}
\affiliation{Instituto de F\'isica de S\~{a}o Carlos, Universidade de S\~{a}o Paulo - 13566-590 S\~{a}o Carlos, SP, Brazil}

\author{S.E. Skipetrov}
\affiliation{Universit\'e Grenoble Alpes, CNRS, LPMMC, 38000 Grenoble, France}

\author{R. Kaiser}
\affiliation{Universit\'e C\^ote d'Azur, CNRS, INPHYNI, France}

\author{R. Bachelard}
\affiliation{Departamento de F\'{\i}sica, Universidade Federal de S\~{a}o Carlos, Rodovia Washington Lu\'{\i}s, km 235 - SP-310, 13565-905 S\~{a}o Carlos, SP, Brazil}
\affiliation{Universit\'e C\^ote d'Azur, CNRS, INPHYNI, France}

\date{\today}

\begin{abstract}
Light propagation in disordered media is particularly complex due to the interference effects and polarization-mixing terms. Two-dimensional systems are particularly interesting because they feature two decoupled scattering channels, with distinct spectral statistics due to the presence of polarization-coupling terms in only one of them. We here show that in one channel, Anderson localization bounds the light transport both longitudinally and transversely. In the second channel, in contrast, diffusive transport takes place
independently of scatterer density. Our results open new possibilities for engineering of light transport in two-dimensional waveguides.
\end{abstract}

\maketitle


\vspace{1cm}


{\em Introduction.---} Controlling light transport in disordered media is an important goal for the engineering of photonic devices~\cite{cao2022harnessing,vynck2023light}. Indeed, interference in disordered systems is central to a broad range of applications~\cite{Sauty2022}, including enhanced light trapping for absorption management in photovoltaics panels~\cite{sheremet2020absorption}, structural color and whiteness control in appearance of materials~\cite{garcia2007photonic,forster2009biomimetic,schertel2019structural}, and transparency engineering in hyperuniform architectures~\cite{maurice1957structure,hart1969light,benedek1971theory,twersky1975transparency,salameh2020origin}. In this context, the role of polarization in two-dimensional (2D) systems remains particularly interesting, since transverse electric (TE) and tranverse magnetic modes (TM) exhibit fundamentally different transport behaviors~\cite{vynck2023light}.

In 2D disordered systems, Anderson localization~\cite{Anderson1958} plays a paramount role for light transport~\cite{john1987strong,de1989transverse,sperling2013direct,vynck2023light} as recent reports demonstrate for dielectric membranes~\cite{riboli2014engineering,riboli2011anderson,garcia2012nonuniversal}, classical waves~\cite{dalichaouch1991microwave,dutta2020optical} and photonic lattices~\cite{schwartz2007transport}. 2D waveguides present the peculiarity of supporting two decoupled scattering channels~\cite{Maximo2015}. The ``scalar'' one, characterized by a dipole orthogonal to the plane (that is, a single polarization) and radiating isotropically in the plane, exhibits the microscopic signatures expected for Anderson localization, such as exponentially localized, long-lived eigenmodes. In contrast, the ``vectorial'' channel has an in-plane dipole with an anisotropic radiation pattern, two accessible polarizations coupled by scattering, and near-field terms.
The corresponding eigenmodes are extended and much shorter-lived than in the scalar channel~\cite{Maximo2015}. These microscopic features must be placed in the context of the debate about Anderson localization of light, where the vectorial nature of electromagnetic waves has been shown to prevent localization~\cite{Skipetrov2014, Maximo2015}. Since then, the study of spectral statistics showed how to restore localization in three dimensions, using either external fields~\cite{Skipetrov2015} or diagonal disorder~\cite{celardo2024localization}. But while light propagation was predicted to exhibit localization features in 3D systems of metallic scatterers~\cite{yamilov2025anderson}, the transport properties in two-dimensional waveguides remain to be investigated~\cite{Laurent2007, Yang2026}.  

In this work, we show that transport of light in the scalar channel of a 2D waveguide filled with point scatterers is dominated by interference effects (Anderson localization), leading to light confinement both longitudinally and transversely. The localization length, which is determined by the scatterer density, sets the length scale for this confinement, and is well approximated by the self-consistent theory of localization. In contrast, the vectorial channel exhibits diffusive-like transport properties, with Ohm's law for the transmission and diffusive spreading of light inside the sample. Hence, point dipoles in 2D waveguides appear to be an ideal platform to control the transport of light in disordered media through its polarization.

\begin{figure}[b]
\centering
\includegraphics[width=1.0\columnwidth]{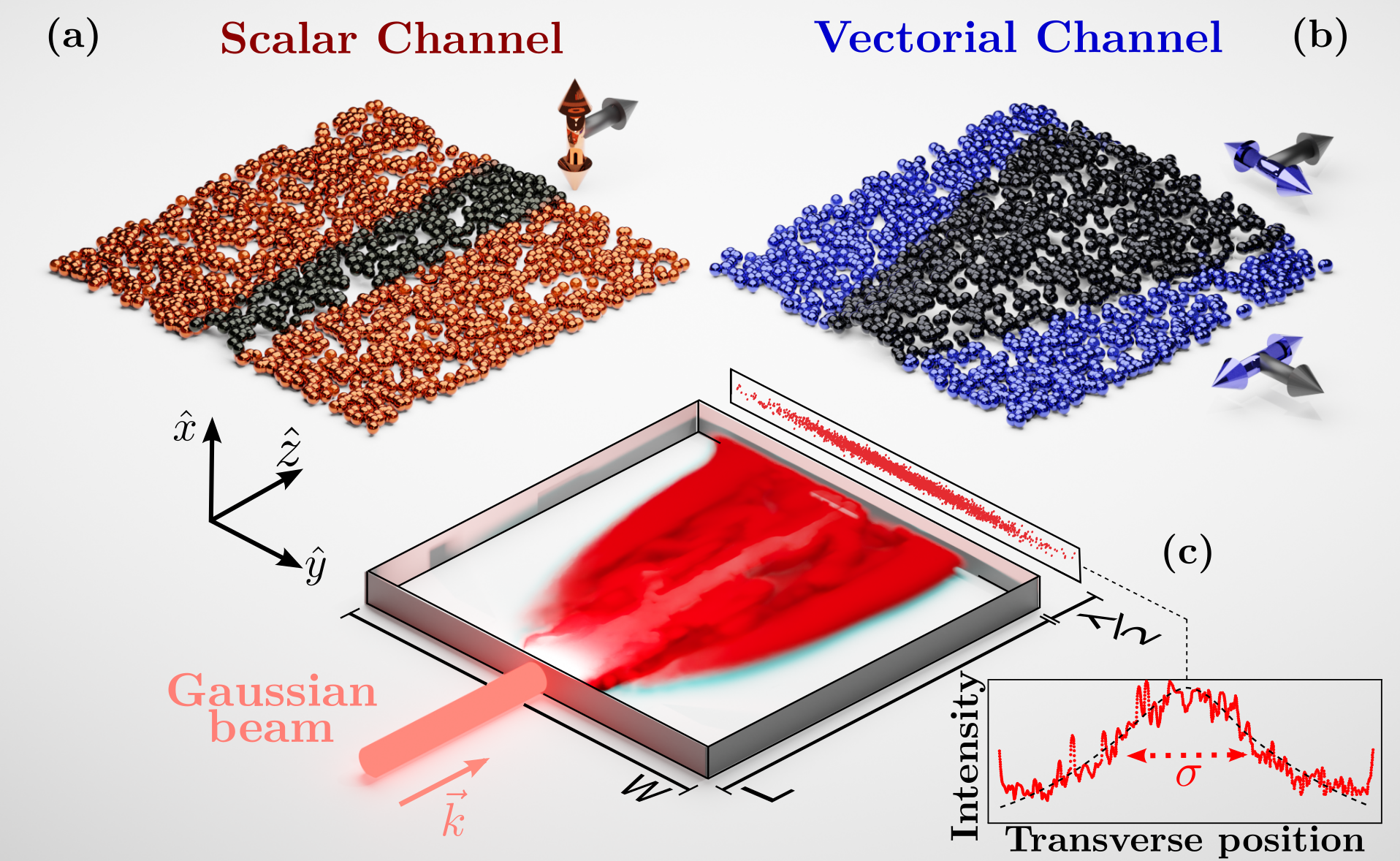}
\caption{(Color online)  (a--b) Scheme of light propagation in (a) the scalar channel, where the incident drive has a polarization orthogonal to the waveguide, and (b) the vectorial channel, with an in-plane polarization. (c) Propagating beam profile in the scalar channel, with its transverse profile in the exit plane shown in the inset.}
\label{fig:scheme}
\end{figure}

{\em Scattering in a 2D ensemble of point-like scatterers.---} Let us consider a set of $N$ two-level atoms at random positions $\mathbf{r}_j = (y_j, z_j)$ in a 2D waveguide where a single transverse optical mode is present. Interaction of atoms with light is treated in the dipole approximation. The atomic sample has a shape of a 2D slab of thickness $L$ along $z$ axis and transverse width $W$ along $y$ axis. In 2D, for dilute samples, two decoupled scattering channels emerge~\cite{Maximo2015} (see Fig.~\ref{fig:scheme}), associated to atomic dipoles oscillating along $x$ axis, with a dipole moment $\beta_j^{(x)}$ for the scatterer $j$, and along $\hat{\mathbf{\epsilon}}_{\pm} = (\hat{z}\pm i\hat{y})/\sqrt{2}$ directions, with dipole moments $\beta_j^{(\pm)}$. In other words, incident light with a linear polarization along $x$ preserves its polarization upon scattering, while light with an in-plane polarization scatters through $\hat{\mathbf{\epsilon}}_{\pm}$ channels, yet with its polarization remaining in $yz$ plane.

In the linear optics regime, the dynamics of the two scattering channels is described by the following coupled dipole equations~\cite{Maximo2015}:
\begin{subequations} \label{eq:cde}
\begin{align}
\frac{d\beta_j^{(x)}}{dt} & = \left(i\Delta-\frac{\Gamma}{2}\right)\beta_j^{(x)}-\frac{id}{2\hbar}E_j^{(x)}
\label{eq:sca}
\\ &- \frac{\Gamma}{2}\sum_{m\neq j}H_0(kr_{jm})\beta_m^{(x)}, \nonumber
\\ \frac{d\beta_j^{(\pm)}}{dt} & = \left(i\Delta-\frac{\Gamma}{4}\right)\beta_j^{(\pm)}-\frac{id}{2\hbar}E_j^{(\pm)}
\label{eq:vec}
\\
&- \frac{\Gamma}{4}\sum_{m\neq j}\bigg[ H_0(kr_{jm})\beta_j^{(\pm)}
+e^{2im\phi_{jm}}H_2(kr_{jm})\beta_j^{(\mp)}\bigg], \nonumber
\end{align} 
\end{subequations}
where $\Gamma$ is the resonance linewidth in the scalar channel, $k=2\pi/\lambda$ is the light wavenumber in the free space, $\Delta$ is the laser detuning, $d$ is the electric-dipole transition matrix element, $\mathbf{E}_j$ is the driving electric field on the $j$-th scatterer, $H_n$ is the Hankel function of the first kind and of order $n$, and $\tan\phi_{jm}=(z_j-z_m)/(y_j-y_m)$. Equation \eqref{eq:vec} for the in-plane dipoles $\beta_j^{(\pm)}$ exhibits terms coupling the polarizations $\hat{\mathbf{\epsilon}}_+$ and $\hat{\mathbf{\epsilon}}_-$, thus the corresponding scattering channel is  referred to as vectorial. The $x$-polarized channel corresponding to Eq.\ \eqref{eq:sca} is the scalar one. According to the spectral analysis, the near-field terms in Eq.\ \eqref{eq:vec} are responsible for preventing Anderson localization of vectorial waves~\cite{Maximo2015}, a phenomenon which nevertheless persists in the scalar channel. 

The transport properties of light are derived from the total electric field:
\begin{align}
E_t^{(x)}(\mathbf{r}) = & E_L^{(x)}(\mathbf{r}) - \frac{i \hbar\Gamma}{2d}\sum_{j=1}^{N}{H_0(|\mathbf{r} - \mathbf{r_j}  |)\beta_{j}^{(x)}},
\\ E_t^{(\pm)}(\mathbf{r}) = & E_L^{(\pm)}(\mathbf{r}) - \frac{i \hbar\Gamma}{4d}\sum_{j=1}^{N}  \bigg[H_0(|\mathbf{r} - \mathbf{r_j}  |)\beta_{j}^{(\pm)} \nonumber
\\ & - e^{ \pm 2 i \phi_j } H_2(|\mathbf{r} -\mathbf{r}_j|)\beta^{(\mp)}\bigg] .
\end{align}
In our simulations, the system is driven by a 2D Gaussian beam propagating along $z$ axis~\cite{leseur2014probing}. In the paraxial approximation, it reads:
\begin{equation}
\frac{\mathbf{E}_L(\mathbf{r})}{E_0}= \frac{w_0}{w(z)} \exp\left[\textit{i}k_0 z - \frac{\textit{i}\psi(z)}{2} - \frac{y^2}{2 w_0^2(1+ \textit{i}\frac{z}{z_R})}   \right] \hat{\epsilon},
\label{eq:Gaussiana}
\end{equation}
with $w_0$ the beam waist, $w(z)=w_0\sqrt{1+(z/z_R)^2}$ the local waist, $z_R = kw_0^2/2$ the Rayleigh range and $\psi(z) = \arctan(z/z_R)$ the Gouy phase. In this work, we set $w_0=12\pi/k$ ($8\pi/k$) for the scalar (vectorial) channel, and the waveguide width $W = 48\pi/k$ ($32\pi/k$), which is a compromise between suppressing finite-size effects ($W \gg w_0$) and remaining in the paraxial approximation ($w_0\gg 2\pi/k$) while keeping the total number of atoms $N$ in the sample in the range accessible for numerical simulations ($N \lesssim 10^4$).

\begin{figure}[b]
\centering
\includegraphics[width=1.0\columnwidth]{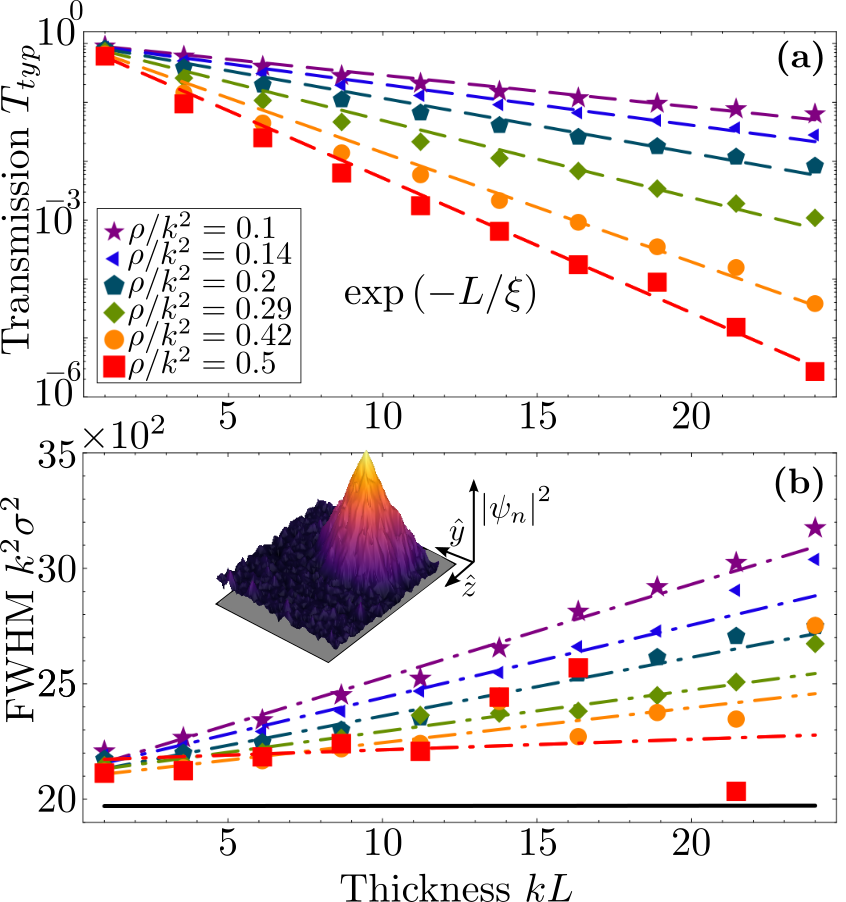}
\caption{(Color online) In the scalar channel, (a) typical transmission, obtained from the geometric average over realizations, $T_\textrm{typ}=\exp(\langle \ln T \rangle)$ and (b) squared transverse width $\sigma^2$ of the transmitted beam, as functions of the normalized thickness of the sample $kL$.  The inset in (b) is an example of a localized quasimode~\cite{Maximo2015}. Lines show exponential and linear fits, respectively, to the microscopic results (symbols). The black line stands for the beam size in absence of the atoms. Simulations realized for samples of width $kW =48 \pi$, the incident beam waist $kw_0 = 12\pi$,
and averaged over $10^4$ random atomic configurations.
}
	\label{fig:signatures_scalar}
\end{figure}

{\em Macroscopic transport laws.---} While scalar waves are expected to be characterized by an exponential decrease of transmission with the sample thickness in the localized regime~\cite{Anderson1958}, observations of this feature have been elusive for electromagnetic waves in 3D. In 1D, the exponential decay has been reported in stacks of dielectric plates~\cite{Berry1996}, yet the importance of precise alignment of plates and the transition to the quasi-1D geometry for imperfect alignment have been pointed out~\cite{schertel2019magnetic}. In 3D, the recent numerical report of localization of electromagnetic waves in disordered media of metallic spheres~\cite{yamilov2023anderson, yamilov2025anderson} will certainly face, in an experimental setup, the caveat of absorption, which also leads to the exponential decay. In 2D, localized modes have been reported in the microwave regime and in the scalar channel~\cite{Laurent2007}, yet transport laws have almost not been investigated~\cite{Yang2026}. This motivates us to explore transport properties of the two polarization channels in details. 

We define the transmission coefficient by integrating the outgoing intensity along $y$ at a distance $\pi/k$ from the end of the sample:
$T = \int_{-y_0}^{y_0}\left|\mathbf{E}_t\right|^2 dy/\int_{-y_0}^{y_0}\left|\mathbf{E}_L\right|^2 dy$, where $y_0=z_\textrm{obs}\tan(\pi/6)$ and 
$z_\textrm{obs}=L+\pi/k$ (that is, a screen at a distance $\pi/k$ from the sample edge, and within an angle $\pm\pi/6$ around the propagation axis). We then average $\ln T$ over multiple random atomic configurations in the sample~\cite{Aharony1996}. Note that different collection angles for the outgoing light yield similar results. As can be observed in Fig.~\ref{fig:signatures_scalar}(a), the transmission in the scalar channel exhibits a clear exponential decay over many decades, which is all the more pronounced as the scatterer number density $\rho$ increases. While this decay is a macroscopic signature of Anderson localization~\cite{Anderson1958} (microscopic Eq.~\eqref{eq:sca} being free from absorption), the faster decay of transmission with increasing densities witnesses the reduction of the localization length (see later discussion).

\begin{figure}[h]
\centering
\includegraphics[width=.98\columnwidth]{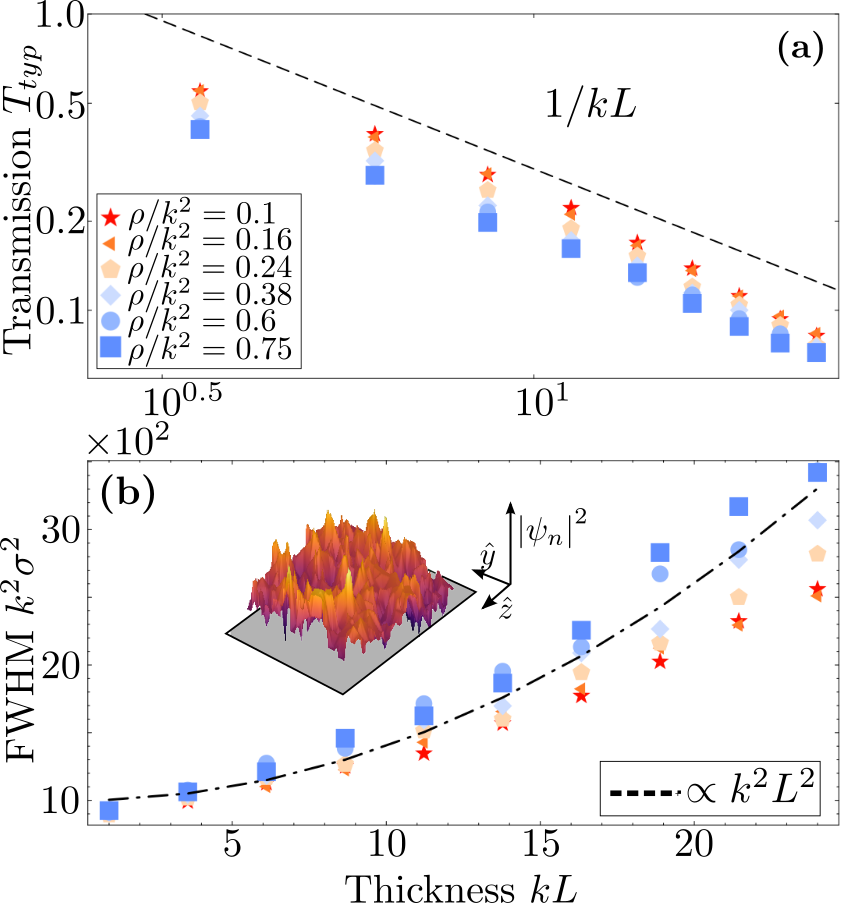}
\caption{(Color online) In the vectorial channel, (a) typical transmission, obtained from a geometric average over realizations, $T_\textrm{typ}=\exp(\langle \ln T \rangle)$ and (b) squared transverse width $\sigma^2$ of the transmitted beam, as functions of normalized sample thickness $kL$.  The inset is an example of an extended quasimode~\cite{Maximo2015}. Lines show (a) the $1/kL$ decay corresponding to the Ohm's law and (b) the quadratic growth $\sigma^2 \propto L^2$ expected in the diffusive regime. Simulations realized for samples of width $kW = 32\pi$, incident beam waist $kw_0 = 8\pi$
and averaged over $10^4$ random atomic configurations.
}
	\label{fig:signatures_vectorial}
\end{figure}

In contrast to the scalar channel, the transmission in the vectorial channel decays as the inverse of sample thickness: $T\propto 1/L$, see Fig.~\ref{fig:signatures_vectorial}(a) 
This power-law behavior corresponds to the Ohm's law for photons, and it is expected for  diffusive transport~\cite{sheng1996introduction, akkermans2007mesoscopic}. The power-law of the decay is the same for different densities of scatterers $\rho$~\cite{rivas2002light}, but the magnitude of $T$ decreases with $\rho$, which can be explained by the corresponding decrease of the photon mean free path. Interestingly, for higher densities the latter decay is slow, which may result from the saturation of transport mean free path at high densities. Thus, the decay of transmission with sample thickness exhibits clear signatures of different scattering regimes in the scalar and vector transport channels.

Another macroscopic feature of light transport is the transverse
spread of the beam propagating through the disordered medium, as illustrated in Fig.~\ref{fig:scheme}(c). Anderson localization is known to drastically reduce this spread \cite{cherroret2010transverse}, with the localization length $\xi$ acting as the spatial scale for the transverse confinement---a phenomenon called transverse localization or confinement~\cite{de1989transverse,sperling2013direct,cherroret2010transverse}. We define the width $\sigma$ of the transmitted beam as the full width at half maximum (FWHM) of the transverse profile of the transmitted intensity. Figure~\ref{fig:signatures_scalar}(b) shows the evolution of $\sigma^2$ with the sample thickness in the scalar channel. Note the close-to-linear growth of $\sigma^2$ with the system size, along with the important reduction of the rate of this growth with the increasing scatterer density $\rho$.
In contrast, in the vectorial channel, the transverse width of the beam grows almost quadratically with $L$ and exhibits much less dependence on the scatterer density $\rho$, see Fig.~\ref{fig:signatures_vectorial}(b). This quadratic growth is expected for diffusive transport~\cite{cherroret2010transverse} whereas the weak dependence on the scatterer density, which cannot be explained based on the independent scatterering approximation, is likely to originate from
a nontrivial density-dependence of the mean free path.

The transport of light resulting from microscopic Eqs.~\eqref{eq:sca} and~\eqref{eq:vec} shows drastic differences between the vectorial and scalar channels, with a weak dependence on scatterer density in the former. In the latter, strong localization features are observed, with the confinement of light which becomes more pronounced with the increasing density, which plays the role of disorder strength. Focusing on the manifestations of localization in the scalar channel, let us now see if they yield the same localization length $\xi$ and compare them with the predictions of the self-consistent theory of localization.

\begin{figure}[t!]
    \centering
	\includegraphics[width=0.98\columnwidth]{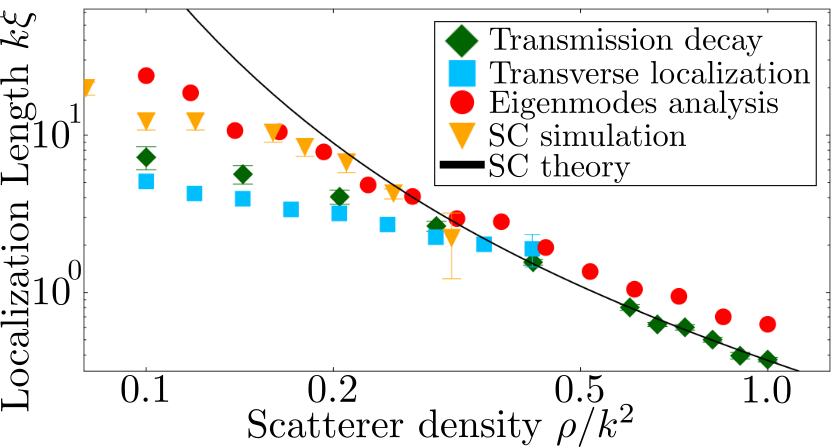}
	\caption{(Color online) Localization length determined from (i) spatial decay of localized eigenvectors $|\psi(\mathbf{r})|^2 \propto\exp(-|\mathbf{r}-\mathbf{r}_\textrm{CM}|/\xi)$ for  $kL=20$, $kW = 48\pi$ (red disks); (ii) transverse width  $\sigma^2 = 2 \xi L$ (blue squares); (iii) transmission decay $\langle \ln T \rangle \propto -L/\xi$ in the microscopic model (green diamonds); (iv) the self-consistent theory of localization, Eq.~\eqref{eq:scthr} (black line); (v) decay of the average transmission $\langle T \rangle \propto\exp(-L/\xi)$ in the self-consistent theory (yellow triangles).}
	\label{fig:xi}
\end{figure}

{\em Localization length: macroscopic, microscopic and self-consistent approaches in the scalar channel.---} The transmission in the localized regime has been predicted to follow an exponential dependence on the sample thickness $L$ set by the localization length: $T\propto \exp(-L/\xi)$~\cite{abrahams1979scaling}. The decay of transmission presented in Fig.~\ref{fig:signatures_scalar}(a) thus allows us to extract $\xi$ using a linear fit to $\ln T_{typ}(L) = \langle \ln T \rangle$. The localization length obtained from this fit is presented in Fig.~\ref{fig:xi} (green diamonds), as a function of the density of scatterers. The decay of $\xi$ with $\rho/k^2$ confirms the role of scatterer density as a disorder parameter.

Transverse localization is also expected to yield a signature of localization length, with $\sigma^2 \propto \xi L$ predicted in 3D in the large-thickness limit, yet with correction terms to account for finite-size effects~\cite{cherroret2010transverse}. We estimate $\xi$ by fitting our simulation results to $\sigma^2= 2\xi L$~\cite{cherroret2010transverse},
which is expected to hold in 2D as well. This yields blue squares in Fig.~\ref{fig:xi}, in good agreement with the results obtained from the transmission decay (green diamonds). Note that the range of densities explored is here limited due to the very low values of transmitted intensity at high densities, which prevents determining the width of transmission profile.

The localization length $\xi$ also manifests itself at the microscopic level, in the exponentially-decaying profile of localized eigenmodes~\cite{Anderson1958}. We extract $\xi$ from a linear fit to the logarithm of the spatial profile of an eigenmode $\psi^{(n)}$ (from Eq.~\eqref{eq:sca} in the steady state and without drive $E_j$): $\ln |\psi^{(n)}(\mathbf{r})|^2 \propto -|\mathbf{r}-\mathbf{r}_\textrm{CM}^{(n)}|/\xi + \textrm{cst} $, where $\mathbf{r}_\textrm{CM}^{(n)}$ is the center of mass of the mode. To account for frequency shifts that appear for increasing density, we selected for each scatterer density $\rho/k^2$ a spectral window  $\omega_n=\Re(\lambda_n) \in [\Delta_\textrm{Loc}-\Gamma/4,\Delta_\textrm{Loc}+\Gamma/4]$, with $\lambda_n$ the associated eigenvalue (see supplemental material for the calculation of $\Delta_{\textrm{Loc}}(\rho)$). The localization length $\xi$ is obtained by selecting the $10\%$ smallest $\xi_n$ in this window, then averaging over disorder realizations. $\xi$ obtained from this microscopic approach is shown in Fig.~\ref{fig:xi} (red disks), and it presents a fair agreement with the macroscopic approaches described above.

Let us now turn to the self-consistent theory of localization---a theory which has been previously shown to agree fairly well with the microscopic transport in disordered ensembles of point scatterers in 2D~\cite{payne2010anderson} and 3D~\cite{Skipetrov2019} for scalar waves. Adopting the approach of Ref.~\cite{cherroret2008}, a local diffusion coefficient $D(\bold{r})$ is defined, from which macroscopic transport properties can be obtained. In 2D, the ensemble-averaged intensity Green’s function $C(\mathbf{r}, \mathbf{r'}$) obeys the following system of self-consistent equations~\cite{payne2010anderson}: 
\begin{subequations}
\begin{align}
    &\left[ - \nabla_{r} \cdot D(\bold{r}) \nabla_{r} \right] C(\bold{r}, \bold{r'}) = \delta(\bold{r} - \bold{r'}),\label{SCIntensityGreen}
    \\
    &\frac{1}{D(\bold{r})} = \frac{1}{D_{B}} + \frac{4}{k l}C(\bold{r}, \bold{r}),\label{SCDiffusionCoeff}
    \\
    &C(\bold{r}, \bold{r'}) - \frac{\pi l}{4}\frac{D(\bold{r})}{D_{B}} (\bold{n} \cdot \nabla) C(\bold{r}, \bold{r'}) = 0 
    \label{eq:sct_C},
\end{align}
\end{subequations}
where $l=k/4\rho$ is the scattering mean free path, $D_{B}$
is the diffusion coefficient in the absence of localization effect, and $\bold{n}$ is a unit vector normal to the sample's boundary and pointing toward its interior.
Equation~\eqref{SCIntensityGreen} describes a stationary diffusion process, Eq.~\eqref{SCDiffusionCoeff} is a self-consistent equation for the position-dependent diffusion coefficient due to interference effects, and Eq.~\eqref{eq:sct_C} represents the boundary condition between the scattering medium and the surrounding free space.

We solve Eqs.~\eqref{SCIntensityGreen}--\eqref{eq:sct_C} numerically by discretizing them on a two-dimensional grid covering the area of our 2D sample~\cite{de2024refroidissement}. We start with $D(\bold{r}) = D_B$, solve Eqs.~\eqref{SCIntensityGreen} and \eqref{eq:sct_C} for $C(\bold{r},\bold{r}')$ and find $C(\bold{r}, \bold{r})$ by regularizing the divergence of $C(\bold{r},\bold{r}')$ at $\bold{r} = \bold{r'}$ with the help of Weierstrass transform (Gaussian filter) with a standard deviation $l$. This allows us to compute the first approximation to $D(\bold{r})$ using Eq.~\eqref{SCDiffusionCoeff}. The procedure is repeated for all points of the grid, and iterated until convergence. We consider that $D(\bold{r})$ has converged to its true value when it changes by less than $10^{-4} \%$ from one iteration to another. To compute the average transmission, we then solve Eq.~\eqref{SCIntensityGreen} with boundary conditions~(\ref{eq:sct_C}) for a Gaussian intensity profile at $z = l$ (i.e., close to the sample boundary illuminated by the incident wave) and compute $\langle T(y) \rangle = -D(y,z)\pdv{C(y,z)}{z}|_{z=L}$~\cite{payne2010anderson}. Note that the solution of  Eq.~\eqref{SCIntensityGreen} is inversely proportional to $D_{B}$, while $D(\bold{r})$ is directly proportional to it. Therefore, the stationary transmission coefficient is independent of $D_{B}$, which exempts us from the need to specify the value of the latter.

\begin{figure}[t!]
\centering
\includegraphics[width=0.98\columnwidth]{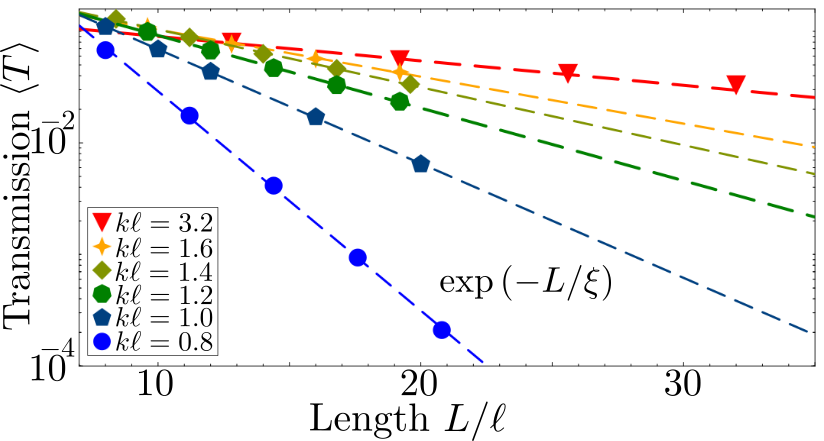}
    \caption{(Color online) 
    Average transmission $\langle T \rangle$ following from the self-consistent theory of localization  as a function of normalized sample thickness $L/\ell$. Here sample width is $kW = 80\pi$, the incident Gaussian beam  waist is $kw_0 = 12\pi$. Dashed lines are fits of the form $\langle T \rangle \propto \exp(-L/\xi)$, used to determine the localization length $\xi$. Only $L/\ell< 30$ has been considered for fits.}
	\label{fig:diffusion}
\end{figure}

The transmission coefficient is obtained from the self-consistent theory by summing $\langle T(y) \rangle$ over $y$.
It exhibits an exponential decay with sample thickness, see Fig.~\ref{fig:diffusion}, similarly to the results of microscopic simulations. The corresponding localization length $\xi$ is again extracted from a linear fit to the logarithm of $\langle T \rangle$. The resulting $\xi$ is in very good agreement with the eigenmode analysis, see orange triangles in Fig.~\ref{fig:xi}. Furthermore, at high densities ($\rho/k^2 \gtrsim 0.2$), $\xi$ also presents a fair agreement with the theoretical prediction for the localization length in 2D~\cite{lee1985disordered,rivas2002light}:
\begin{equation}
    \xi = \frac{k}{4\rho} \exp\left(\frac{\pi k^2}{8\rho}\right),\label{eq:scthr}
\end{equation}
shown in Fig.~\ref{fig:xi} by the solid line. Note that an agreement with this equation has also been observed for Anderson localization of waves in hyperuniform media in the pseudogap frequency range~\cite{Monsarrat2022}.

As can be seen from Fig.~\ref{fig:xi}, discrepancies between localization lengths obtained by different approaches rather occur at low scatterer densities. In particular, this is where the results of microscopic simulations (transmission and transverse profile analysis) deviate most from the predictions of the self-consistent theory. We attribute these deviations to finite-size effects. Indeed, in our microscopic simulations, the maximum scatterer number that can be simulated ($N \sim 10^4$) sets the upper bound on the system area $W \times L$, restricting simulations to relatively small samples. On the contrary, the self-consistent theory allows for considering larger systems, but the use of the Gaussian filter to regularize the divergence of $C(\bold{r}, \bold{r}')$ sets the lower bound for both $W$ and $L$ to approximately $3l$. For consistency, we consider  samples of the same sizes in both microscopic simulations and the self-consistent theory, but these sizes may be only marginally sufficient for the latter to be fully applicable. As for the eigenmode analysis, it may suffer much less from finite-size effects since only the most localized modes are considered and the latter are typically found far from the boundaries.


{\em Conclusion.---} We have demonstrated that the two scattering channels in 2D disordered media exhibit very different transport properties. The scalar channel exhibits halted transport due to interference effect leading to Anderson localization of eigenmodes, with the localization length set by the disorder strength. The self-consistent theory of localization describes this regime rather well, especially at high scatterer densities. In contrast, the vectorial channel exhibits an almost-density-independent diffusive transport obeying the Ohm's law for photons. 
These results highlight the role of polarization in the transport of light in disordered media and the advantage that 2D systems present in terms of control over light propagation due to the existence of two decoupled propagation channels. They open new prospects for controlling optical transport in 2D disordered media, where disorder can become a limiting factor, or on the contrary, a feature to be exploited~\cite{Garcia2010,lucas2026}.



\begin{acknowledgments}
R.B. and J.V.S.F. acknowledge the financial support of the São Paulo Research Foundation (FAPESP) (Grants No. 2023/03300-7, 2022/00209-6, 2025/04954-6, 2023/15110-8 and 2020/14519-1), from the Brazilian CNPq (Conselho Nacional de Desenvolvimento Científico e Tecnológico), Grant No. 403653/2024-0, 403788/2025-0 and 313632/2023-5, and from the French government, through the UCA J.E.D.I. Investments in the Future project managed by the National Research Agency (ANR) with the reference number ANR-15-IDEX-01. RB and RK received support from the project  CAPES-COFECUB (CAPES 88887.711967/2022-00), and RK from the European project ANDLICA, ERC Advanced grant No. 832219, the French National Research Agency (projects PACE-IN (ANR19-QUAN-003), LiLoA (ANR23-CE30-
0035) and QUTISYM (ANR-23-PETQ-0002)).
\end{acknowledgments}

\bibliography{Biblio}

@article{Skipetrov2014,
  title = {Absence of {A}nderson Localization of Light in a Random Ensemble of Point Scatterers},
  author = {Skipetrov, S. E. and Sokolov, I. M.},
  journal = {Phys. Rev. Lett.},
  volume = {112},
  issue = {2},
  pages = {023905},
  numpages = {5},
  year = {2014},
  month = {Jan},
  publisher = {American Physical Society},
  doi = {10.1103/PhysRevLett.112.023905},
  url = {https://link.aps.org/doi/10.1103/PhysRevLett.112.023905}
}

@article{Anderson1958,
  title = {Absence of Diffusion in Certain Random Lattices},
  author = {Anderson, P. W.},
  journal = {Phys. Rev.},
  volume = {109},
  issue = {5},
  pages = {1492--1505},
  numpages = {0},
  year = {1958},
  month = {Mar},
  publisher = {American Physical Society},
  doi = {10.1103/PhysRev.109.1492},
  url = {https://link.aps.org/doi/10.1103/PhysRev.109.1492}
}

@article{Skipetrov2015,
  title = {Magnetic-Field-Driven Localization of Light in a Cold-Atom Gas},
  author = {Skipetrov, S. E. and Sokolov, I. M.},
  journal = {Phys. Rev. Lett.},
  volume = {114},
  issue = {5},
  pages = {053902},
  numpages = {5},
  year = {2015},
  month = {Feb},
  publisher = {American Physical Society},
  doi = {10.1103/PhysRevLett.114.053902},
  url = {https://link.aps.org/doi/10.1103/PhysRevLett.114.053902}
}

@article{Maximo2015,
  title = {Spatial and temporal localization of light in two dimensions},
  author = {M\'aximo, C. E. and Piovella, N. and Courteille, Ph. W. and Kaiser, R. and Bachelard, R.},
  journal = {Phys. Rev. A},
  volume = {92},
  issue = {6},
  pages = {062702},
  numpages = {7},
  year = {2015},
  month = {Dec},
  publisher = {American Physical Society},
  doi = {10.1103/PhysRevA.92.062702},
  url = {https://link.aps.org/doi/10.1103/PhysRevA.92.062702}
}

@article{leseur2014probing,
  title={Probing two-dimensional Anderson localization without statistics},
  author={Leseur, Olivier and Pierrat, Romain and S{\'a}enz, JJ and Carminati, R{\'e}mi},
  journal={Physical Review A},
  volume={90},
  number={5},
  pages={053827},
  year={2014},
  doi={https://doi.org/10.1103/PhysRevA.90.053827},
  url={https://journals.aps.org/pra/abstract/10.1103/PhysRevA.90.053827},
  publisher={APS}
}

@article{schertel2019magnetic,
  title={Magnetic-field effects on one-dimensional Anderson localization of light},
  author={Schertel, Lukas and Irtenkauf, Oliver and Aegerter, Christof M and Maret, Georg and Aubry, Geoffroy J},
  journal={Physical Review A},
  volume={100},
  number={4},
  pages={043818},
  year={2019},
  doi={https://doi.org/10.1103/PhysRevA.100.043818},
  url={https://journals.aps.org/pra/abstract/10.1103/PhysRevA.100.043818},
  publisher={APS}
}

@article{Berry1996,
  title = {Geometric phases from stacks of crystal plates},
  volume = {43},
  ISSN = {0950-0340},
  url = {http://dx.doi.org/10.1080/09500349608232731},
  DOI = {10.1080/09500349608232731},
  number = {1},
  journal = {Journal of Modern Optics},
  publisher = {Informa UK Limited},
  author = {Berry,  M. V. and Klein,  S.},
  year = {1996},
  month = jan,
  pages = {165–180}
}

@article{Laurent2007,
  title = {Localized Modes in a Finite-Size Open Disordered Microwave Cavity},
  author = {Laurent, David and Legrand, Olivier and Sebbah, Patrick and Vanneste, Christian and Mortessagne, Fabrice},
  journal = {Phys. Rev. Lett.},
  volume = {99},
  issue = {25},
  pages = {253902},
  numpages = {4},
  year = {2007},
  month = {Dec},
  publisher = {American Physical Society},
  doi = {10.1103/PhysRevLett.99.253902},
  url = {https://link.aps.org/doi/10.1103/PhysRevLett.99.253902}
}

@article{payne2010anderson,
  title={Anderson localization as position-dependent diffusion in disordered waveguides},
  author={Payne, Ben and Yamilov, Alexey and Skipetrov, Sergey E},
  journal={Physical Review B—Condensed Matter and Materials Physics},
  volume={82},
  number={2},
  pages={024205},
  year={2010},
  url={https://journals.aps.org/prb/abstract/10.1103/PhysRevB.82.024205},
  doi={ https://doi.org/10.1103/PhysRevB.82.024205},
  publisher={APS}
}

@thesis{de2024refroidissement,
  title={Refroidissement laser de l'ytterbium sur la ligne d'intercombinaison pour des exp{\'e}riences sur la localisation de la lumi{\`e}re},
  author={de Melo, {\'A}lvaro Mitchell Galv{\~a}o},
  year={2024},
  school={Universit{\'e} C{\^o}te d'Azur},
  url={https://theses.fr/2024COAZ5033}
}

@article{abrahams1979scaling,
  title={Scaling theory of localization: Absence of quantum diffusion in two dimensions},
  author={Abrahams, Elihu and Anderson, Philip W and Licciardello, Donald C and Ramakrishnan, Tiruppattur V},
  journal={Physical Review Letters},
  volume={42},
  number={10},
  pages={673},
  year={1979},
  url={https://journals.aps.org/prl/abstract/10.1103/PhysRevLett.42.673},
  doi={https://doi.org/10.1103/PhysRevLett.42.673},
  publisher={APS}
}

@article{yamilov2023anderson,
  title={Anderson localization of electromagnetic waves in three dimensions},
  author={Yamilov, Alexey and Skipetrov, Sergey E and Hughes, Tyler W and Minkov, Momchil and Yu, Zongfu and Cao, Hui},
  journal={Nature physics},
  volume={19},
  number={9},
  pages={1308--1313},
  year={2023},
  url={https://www.nature.com/articles/s41567-023-02091-7},
  doi={https://doi.org/10.1038/s41567-023-02091-7},
  publisher={Nature Publishing Group UK London}
}

@article{sperling2013direct,
  title={Direct determination of the transition to localization of light in three dimensions},
  author={Sperling, Tilo and Buehrer, Wolfgang and Aegerter, Christof M and Maret, Georg},
  journal={Nature Photonics},
  volume={7},
  number={1},
  pages={48--52},
  year={2013},
  url={https://www.nature.com/articles/nphoton.2012.313},
  doi={https://doi.org/10.1038/nphoton.2012.313},
  publisher={Nature Publishing Group UK London}
}

@article{de1989transverse,
  title={Transverse localization of light},
  author={De Raedt, Hans and Lagendijk, AD and de Vries, Pedro},
  journal={Physical review letters},
  volume={62},
  number={1},
  pages={47},
  year={1989},
  url={https://journals.aps.org/prl/abstract/10.1103/PhysRevLett.62.47},
  doi={https://doi.org/10.1103/PhysRevLett.62.47},
  publisher={APS}
}

@article{sheng1996introduction,
  title={Introduction to wave scattering, localization, and mesoscopic phenomena},
  author={Sheng, Ping},
  journal={Taylor \& Francis},
  url ={https://www.tandfonline.com/doi/full/10.1080/17455030701219165},
  year={1996}
}

@article{dalichaouch1991microwave,
  title={Microwave localization by two-dimensional random scattering},
  author={Dalichaouch, Rachida and Armstrong, JP and Schultz, Sheldon and Platzman, PM and McCall, SL},
  journal={Nature},
  volume={354},
  number={6348},
  pages={53--55},
  year={1991},
  url={https://www.nature.com/articles/354053a0},
  doi={https://doi.org/10.1038/354053a0},
  publisher={Nature Publishing Group UK London}
}

@article{dutta2020optical,
  title={Optical absorption microscopy of localized atoms at microwave domain: two-dimensional localization based on the projection of three-dimensional localization},
  author={Dutta, Bibhas Kumar and Panchadhyayee, Pradipta and Bayal, Indranil and Das, Nityananda and Mahapatra, Prasanta Kumar},
  journal={Scientific Reports},
  volume={10},
  number={1},
  pages={536},
  year={2020},
  url={https://www.nature.com/articles/s41598-019-57141-z},
  doi={https://doi.org/10.1038/s41598-019-57141-z},
  publisher={Nature Publishing Group UK London}
}

@article{schwartz2007transport,
  title={Transport and Anderson localization in disordered two-dimensional photonic lattices},
  author={Schwartz, Tal and Bartal, Guy and Fishman, Shmuel and Segev, Mordechai},
  journal={Nature},
  volume={446},
  number={7131},
  pages={52--55},
  year={2007},
  url={https://www.nature.com/articles/nature05623},
  doi={ https://doi.org/10.1038/nature05623},
  publisher={Nature Publishing Group UK London}
}

@article{riboli2014engineering,
  title={Engineering of light confinement in strongly scattering disordered media},
  author={Riboli, Francesco and Caselli, Niccol{\`o} and Vignolini, Silvia and Intonti, Francesca and Vynck, Kevin and Barthelemy, Pierre and Gerardino, Annamaria and Balet, Laurent and Li, Lianhe H and Fiore, Andrea and others},
  journal={Nature materials},
  volume={13},
  number={7},
  pages={720--725},
  year={2014},
  url={https://www.nature.com/articles/nmat3966},
  doi={https://doi.org/10.1038/nmat3966},
  publisher={Nature Publishing Group UK London}
}

@article{riboli2011anderson,
  title={Anderson localization of near-visible light in two dimensions},
  author={Riboli, Francesco and Barthelemy, P and Vignolini, Silvia and Intonti, Francesca and De Rossi, A and Combrie, S and Wiersma, DIEDERIK SYBOLT},
  journal={Optics Letters},
  volume={36},
  number={2},
  pages={127--129},
  year={2011},
  url={https://opg.optica.org/ol/abstract.cfm?uri=ol-36-2-127},
  doi={https://doi.org/10.1364/OL.36.000127},
  publisher={Optical Society of America}
}

@article{garcia2012nonuniversal,
  title={Nonuniversal Intensity Correlations in a Two-Dimensional Anderson-Localizing Random Medium},
  author={Garcia, Pedro David and Stobbe, S{\o}ren and S{\"o}llner, Immo and Lodahl, Peter},
  journal={Physical review letters},
  volume={109},
  number={25},
  pages={253902},
  year={2012},
  url={https://journals.aps.org/prl/abstract/10.1103/PhysRevLett.109.253902},
  doi={https://doi.org/10.1103/PhysRevLett.109.253902},
  publisher={APS}
}

@article{vynck2023light,
  title={Light in correlated disordered media},
  author={Vynck, Kevin and Pierrat, Romain and Carminati, R{\'e}mi and Froufe-P{\'e}rez, Luis S and Scheffold, Frank and Sapienza, Riccardo and Vignolini, Silvia and S{\'a}enz, Juan Jos{\'e}},
  journal={Reviews of Modern Physics},
  volume={95},
  number={4},
  pages={045003},
  year={2023},
  url={https://journals.aps.org/rmp/abstract/10.1103/RevModPhys.95.045003},
  doi={https://doi.org/10.1103/RevModPhys.95.045003},
  publisher={APS}
}

@article{celardo2024localization,
  title={Localization of light in three dimensions: A mobility edge in the imaginary axis in non-Hermitian Hamiltonians},
  author={Celardo, Luca Giuseppe and Angeli, Mattia and Mattiotti, Francesco and Kaiser, Robin},
  journal={Europhysics Letters},
  volume={145},
  number={3},
  pages={35002},
  year={2024},
  doi={10.1209/0295-5075/ad222c},
  url={https://iopscience.iop.org/article/10.1209/0295-5075/ad222c},
  publisher={IOP Publishing}
}

@article{yamilov2025anderson,
  title={Anderson transition for light in a three-dimensional random medium},
  author={Yamilov, Alexey and Cao, Hui and Skipetrov, Sergey E},
  journal={Physical review letters},
  volume={134},
  number={4},
  pages={046302},
  year={2025},
  url={https://journals.aps.org/prl/abstract/10.1103/PhysRevLett.134.046302},
  doi={https://doi.org/10.1103/PhysRevLett.134.046302},
  publisher={APS}
}

@article{john1987strong,
  title={Strong localization of photons in certain disordered dielectric superlattices},
  author={John, Sajeev},
  journal={Physical review letters},
  volume={58},
  number={23},
  pages={2486},
  year={1987},
  url={https://journals.aps.org/prl/abstract/10.1103/PhysRevLett.58.2486},
  doi={https://doi.org/10.1103/PhysRevLett.58.2486},
  publisher={APS}
}

@book{akkermans2007mesoscopic,
  title={Mesoscopic physics of electrons and photons},
  author={Akkermans, Eric and Montambaux, Gilles},
  year={2007},
  url={https://books.google.com.br/books?hl=pt-BR&lr=&id=cs7AVel15TAC&oi=fnd&pg=PA2&dq=Mesoscopic+physics+of+electrons+and+photons&ots=FfXftVp4h7&sig=xzftKQNqCOY6Ll96pe2BV3vhq48#v=onepage&q=Mesoscopic%20physics%20of%20electrons%20and%20photons&f=false},
  publisher={Cambridge university press}
}

@article{lee1985disordered,
  title={Disordered electronic systems},
  author={Lee, Patrick A and Ramakrishnan, Tiruppattur V},
  journal={Reviews of modern physics},
  volume={57},
  number={2},
  pages={287},
  year={1985},
  url={https://journals.aps.org/rmp/abstract/10.1103/RevModPhys.57.287},
  doi={https://doi.org/10.1103/RevModPhys.57.287},
  publisher={APS}
}

@thesis{rivas2002light,
  title={Light in strongly scattering semiconductors: diffuse transport and Anderson localization},
  author={Rivas, Jaime G{\'o}mez},
  url={http://dare.uva.nl/search?field1=isni;value1=0000000392020297;docsPerPage=1;startDoc=6},
  year={2002}
}

@misc{lucas2026,
      title={Delocalization transition for light in two dimensions}, 
      author={Sébastien Lucas and Christian Miniatura and Sergey E. Skipetrov},
      year={2026},
      eprint={2604.22919},
      archivePrefix={arXiv},
      primaryClass={cond-mat.dis-nn},
      url={https://arxiv.org/abs/2604.22919}, 
}

@article{cherroret2008,
  title = {Microscopic derivation of self-consistent equations of Anderson localization in a disordered medium of finite size},
  author = {Cherroret, N. and Skipetrov, S. E.},
  journal = {Phys. Rev. E},
  volume = {77},
  issue = {4},
  pages = {046608},
  numpages = {9},
  year = {2008},
  month = {Apr},
  publisher = {American Physical Society},
  doi = {10.1103/PhysRevE.77.046608},
  url = {https://link.aps.org/doi/10.1103/PhysRevE.77.046608}
}

@article{cherroret2010transverse,
  title={Transverse confinement of waves in three-dimensional random media},
  author={Cherroret, N and Skipetrov, SE and Van Tiggelen, BA},
  journal={Physical Review E—Statistical, Nonlinear, and Soft Matter Physics},
  volume={82},
  number={5},
  pages={056603},
  year={2010},
  url={https://journals.aps.org/pre/abstract/10.1103/PhysRevE.82.056603},
  doi={https://doi.org/10.1103/PhysRevE.82.056603},
  publisher={APS}
}

@article{cao2022harnessing,
  title={Harnessing disorder for photonic device applications},
  author={Cao, Hui and Eliezer, Yaniv},
  journal={Applied Physics Reviews},
  volume={9},
  number={1},
  year={2022},
  publisher={AIP Publishing}
}

@article{maurice1957structure,
  title={The structure and transparency of the cornea},
  author={Maurice, David M},
  journal={The Journal of physiology},
  volume={136},
  number={2},
  pages={263},
  year={1957}
}

@article{hart1969light,
  title={Light scattering in the cornea},
  author={Hart, Robert W and Farrell, Richard A},
  journal={Journal of the optical society of America},
  volume={59},
  number={6},
  pages={766--774},
  year={1969},
  publisher={Optical Society of America}
}

@article{benedek1971theory,
  title={Theory of transparency of the eye},
  author={Benedek, GB},
  journal={Applied optics},
  volume={10},
  number={3},
  pages={459--473},
  year={1971},
  url={https://opg.optica.org/ao/fulltext.cfm?uri=ao-10-3-459},
  doi={https://doi.org/10.1364/AO.10.000459},
  publisher={Optical Society of America}
}

@article{twersky1975transparency,
  title={Transparency of pair-correlated, random distributions of small scatterers, with applications to the cornea},
  author={Twersky, Victor},
  journal={Journal of the Optical Society of America},
  volume={65},
  number={5},
  pages={524--530},
  year={1975},
  url={https://opg.optica.org/josa/abstract.cfm?uri=josa-65-5-524},
  doi={https://doi.org/10.1364/JOSA.65.000524},
  publisher={OSA}
}

@article{salameh2020origin,
  title={Origin of transparency in scattering biomimetic collagen materials},
  author={Salameh, Chrystelle and Salviat, Flore and Bessot, Elora and Lama, Mil{\'e}na and Chassot, Jean-Marie and Moulongui, Elodie and Wang, Yan and Robin, Marc and Bardouil, Arnaud and Selmane, Mohamed and others},
  journal={Proceedings of the National Academy of Sciences},
  volume={117},
  number={22},
  pages={11947--11953},
  year={2020},
  url={https://www.pnas.org/doi/abs/10.1073/pnas.2001178117},
  doi={https://doi.org/10.1073/pnas.200117811},
  publisher={National Academy of Sciences}
}

@article{garcia2007photonic,
  title={Photonic glass: a novel random material for light},
  author={Garc{\'\i}a, Pedro David and Sapienza, Riccardo and Blanco, {\'A}lvaro and L{\'o}pez, Cefe},
  journal={Advanced Materials},
  volume={19},
  number={18},
  pages={2597--2602},
  year={2007},
  url={https://www.researchgate.net/publication/220005636_Photonic_Glass_a_Novel_Random_Material_for_Light},
  doi={10.1002/adma.200602426},
  publisher={WILEY-VCH Verlag Weinheim}
}

@article{forster2009biomimetic,
  title={Biomimetic isotropic nanostructures for structural coloration},
  author={Forster, Jason D and Noh, Heeso and Liew, Seng Fatt and Saranathan, Vinodkumar and Schreck, Carl F and Yang, Lin and Park, Jin-Gyu and Prum, Richard O and O'Hern, Corey S and Mochrie, Simon GJ and others},
  journal={arXiv preprint arXiv:0912.4290},
  year={2009},
  url={https://arxiv.org/abs/0912.4290},
  doi={https://doi.org/10.1002/adma.200903693},
}

@article{schertel2019structural,
  title={The structural colors of photonic glasses},
  author={Schertel, Lukas and Siedentop, Lukas and Meijer, Janne-Mieke and Keim, Peter and Aegerter, Christof M and Aubry, Geoffroy J and Maret, Georg},
  journal={Advanced Optical Materials},
  volume={7},
  number={15},
  pages={1900442},
  year={2019},
  publisher={Wiley Online Library},
  url={https://advanced.onlinelibrary.wiley.com/doi/full/10.1002/adom.201900442},
  doi={ https://doi.org/10.1002/adom.201900442},
}

@article{sheremet2020absorption,
  title={Absorption of scalar waves in correlated disordered media and its maximization using stealth hyperuniformity},
  author={Sheremet, Alexandra and Pierrat, Romain and Carminati, R{\'e}mi},
  journal={Physical Review A},
  volume={101},
  number={5},
  pages={053829},
  year={2020},
  publisher={APS},
  url={https://journals.aps.org/pra/abstract/10.1103/PhysRevA.101.053829},
  doi={https://doi.org/10.1103/PhysRevA.101.053829},
}

@article{Skipetrov2019,
  title = {Intensity of Waves Inside a Strongly Disordered Medium},
  author = {Skipetrov, S. E. and Sokolov, I. M.},
  journal = {Phys. Rev. Lett.},
  volume = {123},
  issue = {23},
  pages = {233903},
  numpages = {6},
  year = {2019},
  month = {Dec},
  publisher = {American Physical Society},
  doi = {10.1103/PhysRevLett.123.233903},
  url = {https://link.aps.org/doi/10.1103/PhysRevLett.123.233903}
}

@article{Monsarrat2022,
  title = {Pseudogap and Anderson localization of light in correlated disordered media},
  author = {Monsarrat, R. and Pierrat, R. and Tourin, A. and Goetschy, A.},
  journal = {Phys. Rev. Res.},
  volume = {4},
  issue = {3},
  pages = {033246},
  numpages = {22},
  year = {2022},
  month = {Sep},
  publisher = {American Physical Society},
  doi = {10.1103/PhysRevResearch.4.033246},
  url = {https://link.aps.org/doi/10.1103/PhysRevResearch.4.033246}
}

@article{Sauty2022,
  title = {Localization Effect in Photoelectron Transport Induced by Alloy Disorder in Nitride Semiconductor Compounds},
  author = {Sauty, Myl\`ene and Lopes, Nicolas M. S. and Banon, Jean-Philippe and Lassailly, Yves and Martinelli, Lucio and Alhassan, Abdullah and Chow, Yi Chao and Nakamura, Shuji and Speck, James S. and Weisbuch, Claude and Peretti, Jacques},
  journal = {Phys. Rev. Lett.},
  volume = {129},
  issue = {21},
  pages = {216602},
  numpages = {6},
  year = {2022},
  month = {Nov},
  publisher = {American Physical Society},
  doi = {10.1103/PhysRevLett.129.216602},
  url = {https://link.aps.org/doi/10.1103/PhysRevLett.129.216602}
}

@article{Aharony1996,
  title = {Absence of Self-Averaging and Universal Fluctuations in Random Systems near Critical Points},
  author = {Aharony, Amnon and Harris, A. Brooks},
  journal = {Phys. Rev. Lett.},
  volume = {77},
  issue = {18},
  pages = {3700--3703},
  numpages = {0},
  year = {1996},
  month = {Oct},
  publisher = {American Physical Society},
  doi = {10.1103/PhysRevLett.77.3700},
  url = {https://link.aps.org/doi/10.1103/PhysRevLett.77.3700}
}

@article{Garcia2010,
  title = {Density of states controls Anderson localization in disordered photonic crystal waveguides},
  author = {Garc\'{\i}a, P. D. and Smolka, S. and Stobbe, S. and Lodahl, P.},
  journal = {Phys. Rev. B},
  volume = {82},
  issue = {16},
  pages = {165103},
  numpages = {5},
  year = {2010},
  month = {Oct},
  publisher = {American Physical Society},
  doi = {10.1103/PhysRevB.82.165103},
  url = {https://link.aps.org/doi/10.1103/PhysRevB.82.165103}
}

@article{Yang2026,
  title = {Transverse Anderson localization in disordered optical media with independently controlled scatterer parameters by ion track technology},
  volume = {615},
  ISSN = {0030-4018},
  url = {http://dx.doi.org/10.1016/j.optcom.2026.133280},
  DOI = {10.1016/j.optcom.2026.133280},
  journal = {Optics Communications},
  publisher = {Elsevier BV},
  author = {Yang,  Pengchong and Xu,  Jiawei and Tao,  Lei and Xue,  Haizhou and Cai,  Li and Cheng,  Hongwei and Lyu,  Shuangbao and Huang,  Ran and Duan,  Jinglai},
  year = {2026},
  month = Oct,
  pages = {133280}
}


\onecolumngrid
\newpage
\begin{center}
{\large{ {\bf Supplemental Material for\\
``Polarization-controlled transport of light in a two-dimensional waveguide''}}}

\setcounter{page}{1}

\vskip0.5\baselineskip{J.V.S. Ferreira,{$^{1,2}$} A.M.G de Melo,{$^2$} N.A. Moreira,{$^3$} S. Skipetrov,{$^4$} R. Kaiser,{$^2$} and R. Bachelard{$^{1,2}$}}

\vskip0.5\baselineskip{ {\it $^{1}$Departamento de Física, Universidade Federal de São Carlos,\ Rodovia Washington Luís, km 235 - SP-310, 13565-905 São Carlos,  SP, Brazil}}
\vskip0.5\baselineskip{{\it $^{2}$Universit\'e C\^ote d'Azur, CNRS, INPHYNI, France}}
\vskip0.5\baselineskip{{\it $^{3}$Instituto de F\'isica de S\~{a}o Carlos, Universidade de S\~{a}o Paulo - 13566-590 S\~{a}o Carlos, SP, Brazil}}
\vskip0.5\baselineskip{{\it $^{4}$University Grenoble Alpes, CNRS, LPMMC, 38000 Grenoble, France}}

\end{center}

\appendix

\section{Frequency shift of localized modes}

A spectral analysis reveals that the region of the spectrum where the localized collective modes are found is shifted from the atomic resonance~\cite{Maximo2015}. Using the coupled dipole model, we realize a mapping of the mean energy of the localized modes as a function of the scatterer density, $\Delta_{\mathrm{LOC}} (\rho)$. 

To this end we implement the following numerical procedure. A cloud of $N$ scatterers randomly distributed
in a disk is generated.
The scattering matrix associated to Eq.~\eqref{eq:sca} is given by 
\begin{equation}
 M_{jm}=\delta_{jm}\left(i\Delta-\frac{\Gamma}{2}\right)-(1-\delta_{jm})\frac{\Gamma}{2}H_0(kr_{jm}).
\end{equation}
Its normalized eigenvectors $\psi_n$ for $\Delta =0$ 
are then computed.
As illustrated in Fig.~\ref{fig:delta_loc}(a), an exponential fit is then performed:
\begin{equation}
\ln |\psi_n^2(\mathbf{r} - \mathbf{r}_\textrm{CM})| =  a -\frac{|\mathbf{r} - \mathbf{r_{CM}}|}{\xi_n},\label{eq:psi}
\end{equation}
where $\mathbf{r}_\textrm{CM}=\sum_j \mathbf{r}_j |\psi_{n,j}| $ is the center of mass of the mode, $\psi_{n,j}$ being the $j$-th component of the $n$-th eigenmode. The localization length $\xi_n$ of this mode is extracted from the fit, $a$ being the other fitting parameter. A criterion on the coefficient of determination of the fit, $R^2 \geq 0.8 $, is applied to identify localized modes, as well as $| \mathbf{r_{CM}}|/R < 0.8$ in order to minimize boundary effects. Eigenvalues $\lambda_n = i\omega_n - \gamma_n/2$ are accumulated from many realizations, each with their own
frequency $\omega_n$ and decay rate $\gamma_n/2$.

\begin{figure}[h]
\centering
\includegraphics[width=1.0\columnwidth]{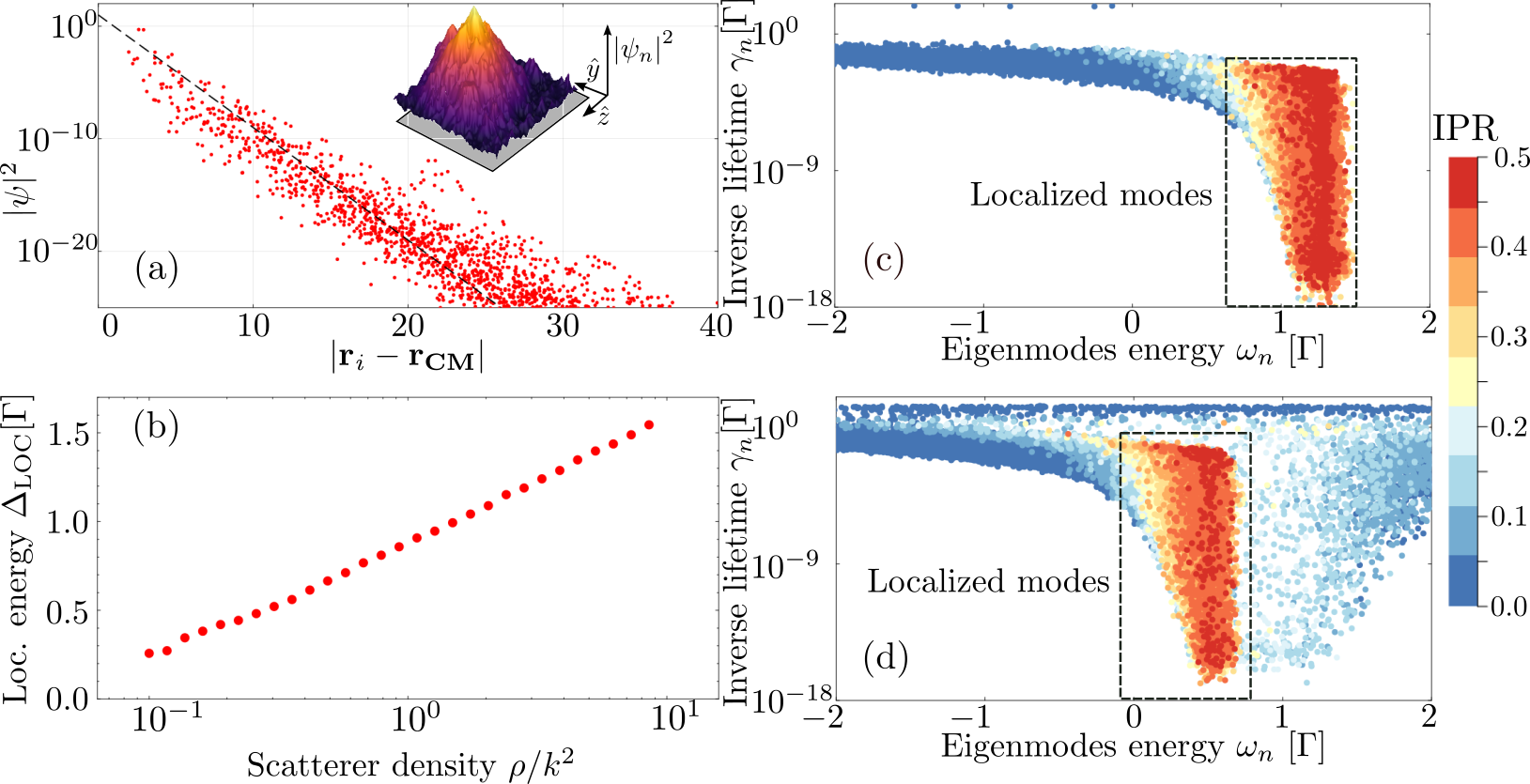}
\caption{(Color online) (a) Spatial profile of a localized eigenmode, the dashed line is a fit of the form~\eqref{eq:psi} and the inset corresponds to its 2D profile. (b)
Frequency shift of the localized modes $\Delta_{\text{LOC}}$ as function of the scatterer density $\rho/k^2$. Simulations realized for
$N = 6000$ scatterers randomly distributed in a disk with a uniform average number density.(c-d) Eigenspectra in the complex plane ($\gamma_n$,$\omega_n$), with the inverse participation ratio as a color code for $\rho/k^2 = 0.5$ and $3$, respectively, where  $\mathrm{IPR}_n = \left(\sum_{j=1}^{N} |\psi_{n,j}|^4\right) \left(\sum_{j=1}^{N} |\psi_{n,j}|^2\right)^{-2}$. }
	\label{fig:delta_loc}
\end{figure}

A complete eigenspectrum is shown in Fig.~\ref{fig:delta_loc}(c-d). From this spectrum the $10\%$ eigenmodes with the lowest $\xi_n$ are selected, and their average frequency shift is used to define $\Delta_\textrm{LOC}$. The procedure is used for different scatterer densities, see Fig.~\ref{fig:delta_loc}(b). The localization length $\xi$ for a given density is then obtained by averaging the set of $\xi_n$ of all localized modes in the frequency window $\omega_n \in [\Delta_\textrm{LOC} - \Gamma/4,\Delta_\textrm{LOC} +\Gamma/4]$, resulting in the eigenmode analysis presented in Fig.~\ref{fig:xi}.    

\end{document}